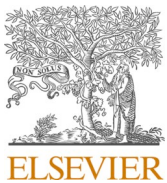
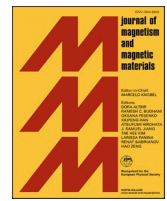

# Effect of reduced local lattice disorder on the magnetic properties of B-site substituted La$_{0.8}$Sr$_{0.2}$MnO$_3$

Sagar Ghorai [a,*], Sergey A. Ivanov [a,b], Ridha Skini [a], Petter Ström [c], Peter Svedlindh [a]

[a] *Department of Materials Science and Engineering, Uppsala University, Box 35, SE-751 03 Uppsala, Sweden*
[b] *Semenov Institute of Chemical Physics, Kosygina Street, 4, Moscow 119991, Russia*
[c] *Applied Nuclear Physics, Department of Physics and Astronomy, Uppsala University, Box 516, SE-751 20 Uppsala, Sweden*

ABSTRACT

Disorder induced by chemical inhomogeneity and Jahn-Teller (JT) distortions is often observed in mixed valence perovskite manganites. The main reasons for the evolution of this disorder are connected with the cationic size differences and the ratio between JT active and non-JT active ions. The quenched disorder leads to a spin-cluster state above the magnetic transition temperature. The effect of Cu, a B-site substitution in the La$_{0.8}$Sr$_{0.2}$MnO$_3$ compound, on the disordered phase has been addressed here. X-ray powder diffraction reveals rhombohedral (R-3c) structures for both the parent and B-site substituted compound with negligible change of lattice volume. The chemical compositions of the two compounds were verified by ion beam analysis technique. With the change of electronic bandwidth, the magnetic phase transition temperature has been tuned towards room temperature (318 K), an important requirement for room temperature magnetic refrigeration. However, a small decrease of the isothermal entropy change was observed with Cu-substitution, related to the decrease of the saturation magnetization.

## 1. Introduction

Mixed valence perovskite manganite oxides (A$_x$B$_{1-x}$MnO$_3$) have attained much attention owing to magnetic disorder driven by competing magnetic interactions and coupling between charge, spin, lattice and orbital degrees of freedom [1]. Mostly, they exhibit two types of exchange interactions; ferromagnetic (FM) double-exchange interaction via Mn$^{3+}$- O$^{2-}$- Mn$^{4+}$ and antiferromagnetic (AFM) super-exchange interaction via Mn$^{3+}$- O$^{2-}$- Mn$^{3+}$ (or Mn$^{4+}$- O$^{2-}$- Mn$^{4+}$). Depending on the Mn$^{3+}$/Mn$^{4+}$ ratio there will be a varying degree of competing FM-AFM interactions. The competing interactions can cause magnetic disorder in manganites, which is often revealed by a Griffiths phase (GP) singularity [2]. In the original work of Griffiths [3], a randomly diluted Ising ferromagnetic system was considered with only a fraction of the lattice-sites occupied with nearest-neighbour interacting Ising spins. If the lattice system is considered as a state $\psi$, and lattice-sites with and without Ising spins are described as $v(\psi)$ and $s(\psi)$, respectively, then the probability of the lattice system can be written as,

$$P(\psi) = P(v) + P(s) = 1$$

For an undiluted or homogeneous ferromagnetic system, $P(s) = 0$ [2–5]. For $P(v) < 1$, above a certain value (percolation threshold), long-range ferromagnetic order is complete at a probabilistic transition temperature $T_C(P(v))$, which is less than the transition temperature of an undiluted system [3]. In case of $P(v) < 1$, ferromagnetic order begins to develop below the Griffiths temperature ($T_G$) as finite size ferromagnetically ordered spin-clusters. The temperature region between $T_G$ and $T_C$ is defined as the GP-region [4]. Obviously, the width of the GP-region region depends on $P(v)$, but in systems with competing FM and AFM interactions it will also depend on the relative amount and strength of these interactions. Thus, in perovskite manganites with competing exchange interactions, the Mn$^{3+}$/Mn$^{4+}$ ratio can tune the width of the GP-region.

Previously, the evolution of the GP-region in manganites has been studied for different A-site substitutions [5–11,12]. However, in most cases A-site substitution introduces a change of lattice volume or even a change in crystal structure, which does not concur with the original model of Griffiths [3]. Thus, A-site substitution often introduces additional changes in the system that can mask the effect of disorder on the evolution of the GP phase. There are only a few reports, [2,13] which describe the dependence of the GP phase on B-site substitution and the reason for an increasing or decreasing width of the GP region in B-site substituted manganites is not clear. The strength and the relative amount of FM and AFM interactions between B-site atoms together with local lattice distortions due to Jahn-Teller active ions govern the formation and evolution of ferromagnetic clusters above $T_C$.

In this work we have substituted the B-site of $La_{0.8}Sr_{0.2}MnO_3$ with magnetic Cu-atoms. With the support of electronic structure analysis, we have characterized the magnetic interactions between B-site atoms and their effect on the formation of the GP phase. As manganites can be tuned to act as a magnetic refrigerant near room temperature, [14,15] the effect of B-site substitution on the magnetocaloric effect has also been studied in this work. A comparatively high value of isothermal entropy change (see Table 2) over a wide temperature span, makes this substitution interesting for room temperature magnetic refrigeration applications.

## 2. Experimental details

The $La_{0.8}Sr_{0.2}MnO_3$ (**LS**) and $La_{0.8}Sr_{0.2}Mn_{0.9}Cu_{0.1}O_3$ (**LSC**) compounds were prepared by solid-state reaction. Stoichiometric amounts of $La_2O_3$, $Sr_2O_3$, $MnCO_3$ and CuO powders were mixed together and calcinated at 1473 K for 24 h in Ar-atmosphere. The samples were characterized using X-ray powder diffraction (XRPD) at 295 K by using a Bruker D8 Advance diffractometer with Cu-K$_\alpha$ radiation and an angle step size of 0.021°. The elemental analysis of the samples was performed by time-of-flight elastic recoil detection analysis (ToF-ERDA) [16] with 36 MeV $^{127}I^{8+}$. The incidence angle of the ion-beam was 23°±1° with respect to the sample surface, and recoils were detected at 45°. Simultaneous Rutherford backscattering spectrometry (RBS) and particle induced X-ray emission (PIXE) with a 2 MeV $^4He^+$ beam and detectors at 170° (RBS) and 135° (PIXE) were also applied. X-ray photoelectron spectroscopy (XPS) was used to analyse the oxidation states and valence band spectra of the samples. The XPS spectra were collected by using a "PHI Quantera II" system with an Al-K$\alpha$ X-ray source and a hemispherical electron energy analyser having a pass energy of 26.00 eV. Presputtering with Ar-ions of 200 eV for 30 s was done on the samples before collecting the XPS spectra in order to remove surface impurities without affecting the sample's properties. A Quantum Design MPMS XL system was used to measure the magnetic properties in the temperature range from 390 K to 5 K with a maximum field of 5 T.

## 3. Results and discussion

### 3.1. Crystal structure

The analysis of the XRPD spectra (Fig. 1(a) and (b)) with the Fullprof program [17] reveals a single phase rhombohedral structure for the two compounds. During XRPD analysis several structural models, orthorhombic, monoclinic, rhombohedral, etc. were fitted and the best fit of the structural model was observed for the rhombohedral structure with a space group R-3c. The structural parameters are listed in Table 1. The absence extra peaks and similar lattice parameters in the Cu-substituted compound, confirm that the Cu-atoms have substituted the Mn-site (6e-site) of the lattice.

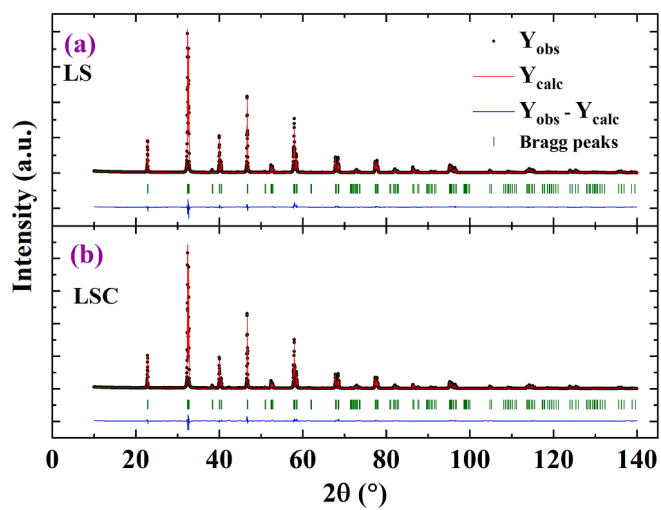

**Fig. 1.** X-Ray powder diffraction patterns of **(a)** LS and **(b)** LSC samples.

The crystal structure can also be described with the Goldschmidt tolerance factor [18] ($t_G = \frac{r_A + r_O}{\sqrt{2}(r_B + r_O)}$, where $r_A$, $r_B$ and $r_O$ are the ionic radii of A, B and oxygen ions, respectively). For the LS and LSC compounds the values of $t_G$ are very close (0.921 and 0.923, respectively) even if the mixed valence of Cu (+2 and +3 oxidation states) is considered (from XPS analysis, described later). The $t_G$ values are calculated using the ionic radii (calculated by R. D. Shannon[19]) of the $La^{+3}$, $Sr^{2+}$, $Mn^{+3}$, $Mn^{+4}$, $Cu^{+2}$, $Cu^{+3}$ and $O^{-2}$ ions as 1.216, 1.31, 0.645, 0.53, 0.73, 0.54 and 1.4 Å, respectively. The lattice distortion in the compounds can be understood more clearly in terms of the deviation from cubic symmetry defined as, $q = \frac{c/2\sqrt{3}}{a/\sqrt{2}}$ [20,21]. The almost same values of $q$ for the two compounds are listed in Table 1. This minimal difference in the crystal structures of the two compounds provides the opportunity to study the effect of disorder and GP-evolution without influence from a crystal structure change as mentioned in the introduction.

### 3.2. Electronic structure

The magnetic properties of the LS and LSC compounds are strongly dependent on the amount of Mn and Cu mixed valence states. The exchange interactions in the compounds, e.g. ferromagnetic $Mn^{+3}$-$O^{-2}$-$Mn^{+4}$, antiferromagnetic $Mn^{+3}$-$O^{-2}$-$Mn^{+3}$ (or $Mn^{+4}$-$O^{-2}$-$Mn^{+4}$), antiferromagnetic $Mn^{+3}$-$O^{-2}$-$Cu^{+2}$ (for LSC compound), depend on the relative amounts of the different oxidation states of the magnetic Mn and Cu-ions. In order to, carefully investigate the oxidation states of these two elements, XPS analysis has been performed. Mainly, two oxidation states for both Mn (+3 and +4) and Cu (+2 and +3) were observed from the XPS core level spectra of Mn *2p* (Fig. 2(*a*)) and Cu *2p* (Fig. 2(*b*)) states. Owing to the spin orbital splitting of Mn $2p_{3/2}$ and Mn $2p_{1/2}$ states two distinct peaks near ~641 eV and ~653 eV were observed for the both compounds (cf. Fig. 2(*a*)). However, in the LSC compound, a broad feature near ~647 eV was also observed and this feature is not a satellite peak for the $Mn^{+2}$ state [23]. M.S. Kim *et al.* [24] identified this feature as a mixed state of $Mn^{3+}$ and $Mn^{4+}$. They also observed that with increasing Cu-substitution in the $La_{0.7}Sr_{0.3}MnO_3$ system this mixed state feature shifts towards higher binding energy as a result of an increasing relative amount of $Mn^{4+}$ ions in the compound. Thus, for the LSC compound the increased amount of $Mn^{4+}$-ions could explain the observation of this broad feature in the Mn *2p* core level spectrum. However, owing to the presence of this ~647 eV feature, any type of quantitative analysis of the Mn 2p peak is difficult. The Mn *3s* spectra is more precise for identification of the Mn oxidation states. The Mn 3s spectra have two distinct peaks originating from the parallel and anti-parallel coupling between the Mn 3s core holes and the Mn 3d electrons [25]. There is a linear relationship between the magnitude of the Mn 3s exchange splitting ($\Delta E$) and the spin of the 3d electrons; $\Delta E \propto (2S+1)$ [26]. As references for the Mn$^{+3}$ and Mn$^{+4}$ states, the exchange splitting of the Mn$_2$O$_3$ and MnO$_2$ compounds have been considered (Fig. 2(C)). Also, Beyreuther *et al.* derived a simple relationship between the Mn-valance state ($v$) and $\Delta E$, as,

$$v = 9.67 - 1.27 \times \Delta E \quad (1)$$

**Table 1**
Structural results.

| Compound | | LS | LSC |
|---|---|---|---|
| Space Group | | R-3c | R-3c |
| Lattice parameters (Å) | a | 5.52240(4) | 5.52280(4) |
| | c | 13.33689(12) | 13.3630(11) |
| q | | 0.986 | 0.988 |
| Mn-O Bond lengths (Å) | | 1.9642(7) | 1.9664(7) |
| Mn-O-Mn Bond angle (°) | | 163.90(16) | 162.99(17) |
| Rietveld Refinement Parameters[22] for XRPD | $R_P$ | 7.32 | 6.19 |
| | $R_{WP}$ | 9.77 | 8.31 |
| | $R_B$ | 5.27 | 4.90 |

**Table 2**
Chemical and magnetic results.

| Compound | | LS | LSC |
|---|---|---|---|
| Atomic % from TOF-ERDA | Mn | 22.2(12) | 20.6(11) |
| | Cu | – | 2.0(2) |
| | O | 59.3(33) | 58.3(31) |
| $T_C$(K) | | 330(2) | 318(2) |
| $M_S$(Am$^2$/kg) | | 90.99 | 81.40 |
| $-\Delta S_M^{max}$ (J/kg-K) at 5T | | 5.18 | 4.53 |
| RCP (J/kg) at 5T | | 245 | 230 |
| GP% | $\mu_0H = 0.01T$ | 15% | – |
| | $\mu_0H = 0.05T$ | 13% | – |
| | $\mu_0H = 0.1T$ | 12% | – |

Using the $\Delta E$ values of the reference samples in Eq. (1) and comparing it with the value of LS compound, the value of Mn$^{3+}$/Mn$^{4+}$ for the LS compound was calculated as 2.91(9).

However, for the LSC compound the Mn 3s state is coupled with the Cu 3p state, thus the above-mentioned quantitative analysis cannot be performed for the LSC compound. For the determination of the Mn valence state in the LSC compound, the total chemical formula has been considered, as,

$$La_{0.8}^{+3}Sr_{0.2}^{+2}Mn_x^{+3}Mn_{(1-p-x)}^{+4}Cu_y^{+2}Cu_{(p-y)}^{+3}O_3^{-2}$$

where $x$ and $y$ correspond to the amount of Mn$^{3+}$ and Cu$^{2+}$ ions, respectively, and $p$ to the total amount of Cu ions. From charge neutrality (neglecting any oxygen deficiency) we have,

$$x + y + p = 0.8 \quad (2)$$

Here, $p$ is 0 and 0.1 for the LS and LSC compounds, respectively. From the fitting of the Cu 2p$_{3/2}$ peak (Fig. 2(b)) of the LSC compound, the value of $y$ is calculated as 0.075(25). Using the value $y$ in *Equation (2)*, the value of Mn$^{3+}$/Mn$^{4+}$ for the LSC compound was calculated as 2.3(3). Thus, with Cu-substitution, the Mn$^{3+}$/Mn$^{4+}$ ratio decreases from 2.91(9) to 2.3(3), indicating an increased amount of Mn$^{4+}$-ions in the LSC compound. Considering the error in the XPS results, we have to keep in mind that the calculated values are only approximate estimates, made to understand the increment of Mn$^{4+}$-ions with Cu-substitution. As we will see later (in the magnetic part), the increment of Mn$^{4+}$ ions, is directly proportional to the decrease of the local lattice distortion caused by the Jahn-Teller effect in the compound.

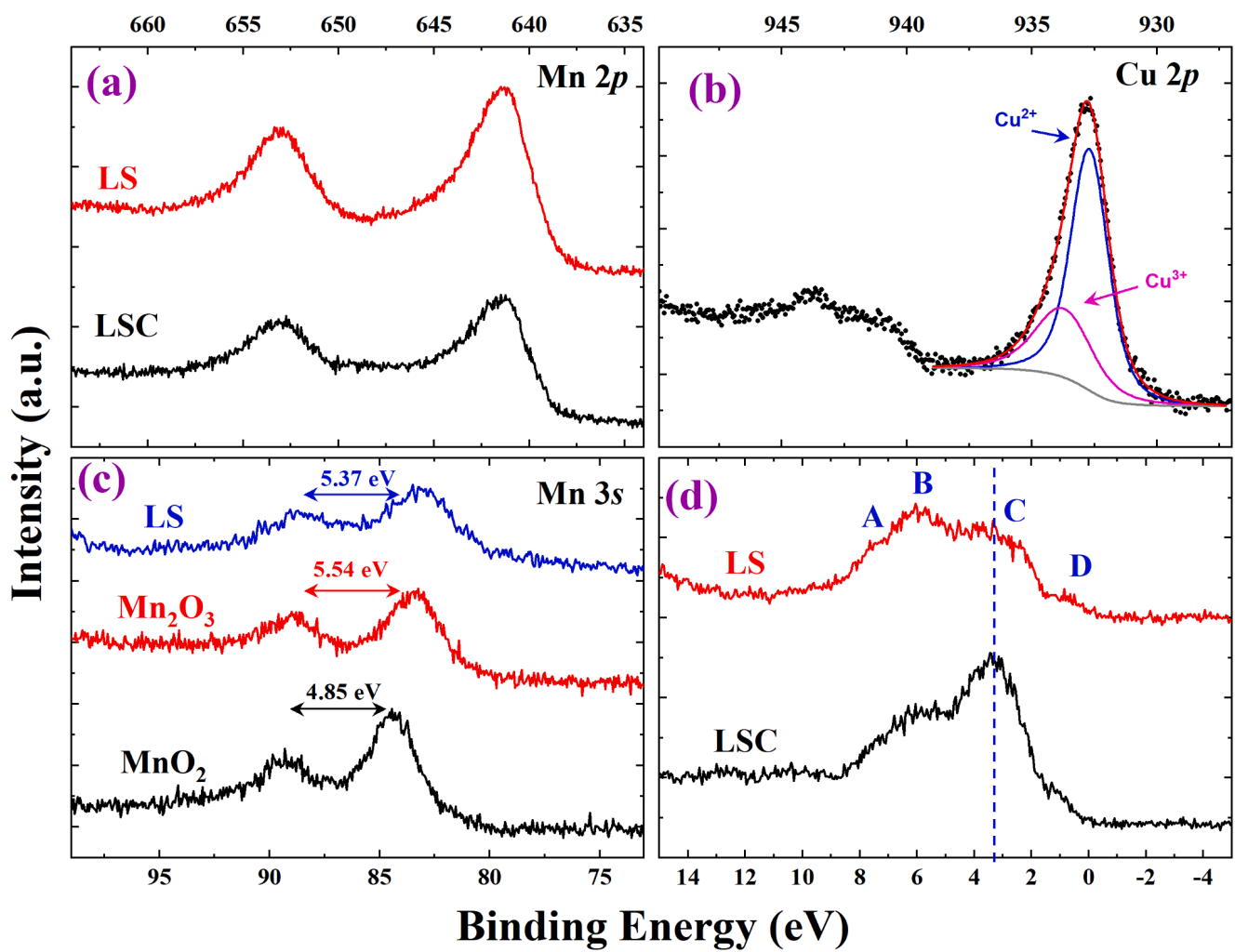

**Fig. 2.** XPS spectra of **(a)** Mn2p state of LS and LSC compounds, **(b)** Cu-2p state of the LSC-compound, **(c)** Mn3s states of LS-and reference compounds and **(d)** valence band spectra of LS and LSC compounds.

The valence band spectra for the LS and LSC compounds are shown in Fig. 2(d). In general, the valence band spectrum of a perovskite manganite has four binding energy contributions A, B, C and D, as indicated in Fig. 2(d), where A corresponds to the O 2p-Mn 3d $t_{2g}$ hybridized state, B to the nonbonding O 2p state, C to the Mn 3d $t_{2g}$ state and D to the Mn 3d $e_g$ state [26–29]. In the LSC compound there is a contribution from the Cu 3d orbital in the binding energy range 2–4 eV [24], which is revealed by the increase of density of states for these energies. Also, from theoretical calculations it is known that only the $e_g$ state of Cu 3d will contribute to the density of states below the Fermi-level [24], which is an indication of the strong coupling between the Mn 3d $t_{2g}$ and Cu 3d $e_g$ orbitals and the nature of this coupling (FM or AFM) will to some extent control the magnetic properties of the LSC compound.

### 3.3. Ion beam analysis

Raw ToF-ERDA data are shown in Fig. 3(a) and (b). In addition to the expected elemental contents, impurity signals due to approximately 0.5–1.5 at. % of H and C were detected on both samples. For the LSC sample, a faint signal due to Si or Al contamination was also detected, part of which may be attributed to the beam grazing the Al sample holder. Depth profiling of the ToF-ERDA data with Potku [30], including all detected signals, and integration from depth $1.5 \times 10^{17}$ at/cm$^2$ to $1.5 \times 10^{18}$ at/cm$^2$ yielded an estimation of the sample composition. The obtained RBS data indicated concentration gradients for La and Sr near the sample surface, making fitting of the relative concentrations ambiguous. Further, a heavy impurity at concentration $\lesssim 0.2$ at. % was detected, identified as Pb from the PIXE spectra shown in Fig. 3(c).

For Mn, the number of counts in the region of the ToF-ERDA spectrum from which data was considered is between 1000 and 2500 for the two samples, yielding a relative statistical uncertainty of 2–3%. The corresponding number for Sr and Cu is 4–5%, while background counts amount to approximately 5–8% of the collected data from these elements. Further, an uncertainty in relative detection efficiency between Mn or Cu and O contributes an error up to 5%. The atomic fractions of Mn, O and Cu obtained from ToF-ERDA are listed in Table 2 and they are comparable with the expected values. Varying the integration depth over a range up to a maximum of $2 \times 10^{18}$ at/cm$^2$, to ascertain the effect of near-surface concentration gradients, yields a variation of the Mn, O and Cu concentrations within the given error margins.

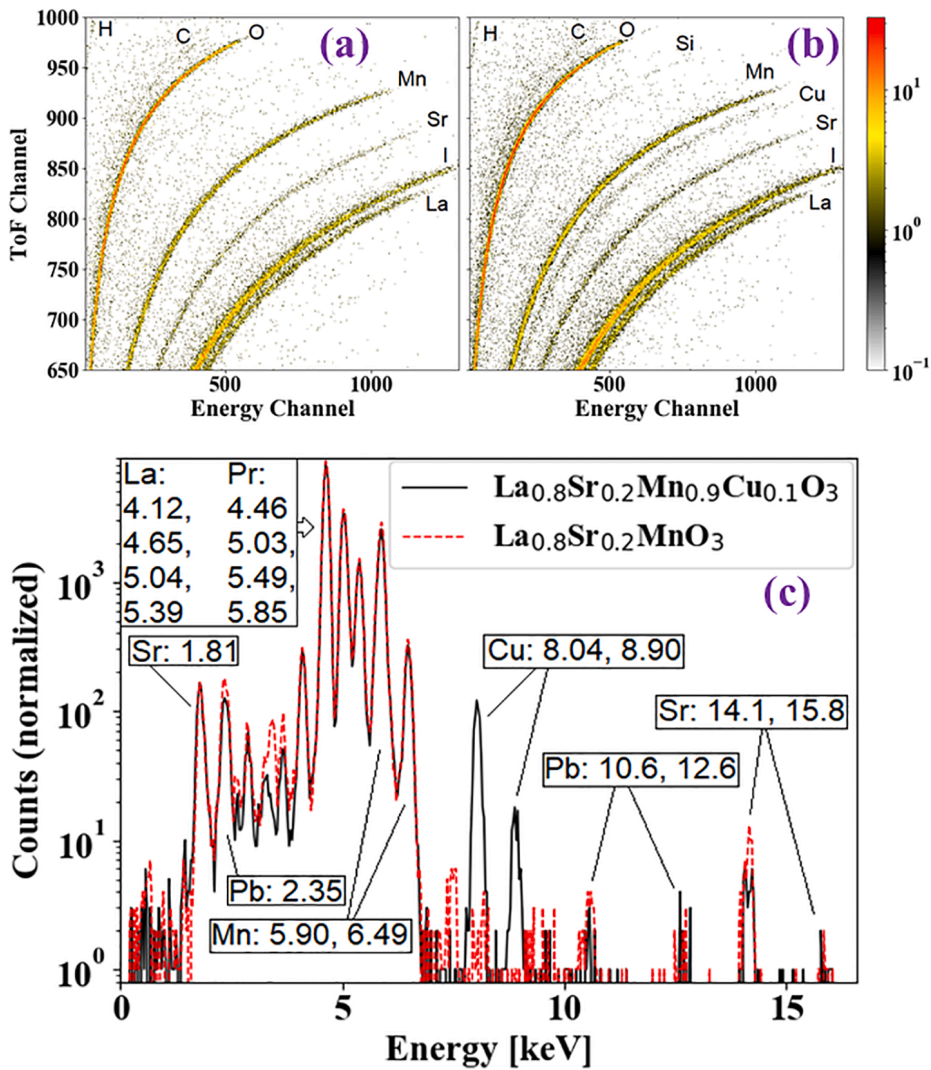

**Fig. 3.** Raw ToF-ERDA data for samples (a) LS and (b) LSC. (c) PIXE spectra obtained on LS and LSC samples, with detected elements indicated. The peaks observed between the Pb Mα-line at 2.35 keV and the set of L-lines from La and Pb are a combination of other Pb M-lines and escape peaks due to excitations in the Si drift detector.

## 3.4. Magnetic and magnetocaloric properties

A second order paramagnetic to ferromagnetic (PM-FM) transition was observed (using Banerjee Criterion[31–34]) in both compounds. However, the transition temperature ($T_C$) decreased with Cu-substitution (from 330 K to 318 K, cf. Fig. 4(a)). Since double-exchange interaction is responsible for the ferromagnetism in the two compounds, one expects that $T_C$ should be governed by the electronic bandwidth ($W$),[35] which is defined as,

$$W \propto \frac{\cos(1/2(\pi - \langle Mn - O - Mn \rangle))}{d_{Mn-O}^{3.5}} \quad (3)$$

where $\langle Mn-O-Mn \rangle$ and $d_{Mn-O}$ are the bond angle and bond length, respectively. From Table 1, a decrease of $W$ is observed with Cu-substitution, which is in accord with the observed decrease of $T_C$, and in good agreement with previously reported results for manganites.[35] Moreover, Cu-substitution introduces a new AFM, $Mn^{3+}$–$O^{2-}$–$Cu^{2+}$ superexchange interaction[36] in the compound and also reduces the saturation magnetization (cf. Fig. 4(b)).

A Griffiths phase like behaviour was observed in the LS compound (Fig. 4(c)), implying that ferromagnetically ordered clusters are formed in the paramagnetic region below the Griffiths temperature $T_G$. The GP, which is suppressed in the LSC compound, can be characterized by the temperature dependence of the magnetic susceptibility according to[2]

$$\chi^{-1} \propto (T - T_C^R)^{1-\lambda} \quad (4)$$

where $0 \leq \lambda < 1$ and $T_C^R$ is the critical temperature where the susceptibility tends to diverge. The different transition temperatures $T_C$, $T_G$, $T_C^R$ are determined following the method discussed by A.K. Pramanik et al. [2] In Fig. 4(c), the red line indicates the fitting of the inverse susceptibility of the LS compound below $T_G$ with **Equation (4)**. The temperature range of the GP is described as,[2]

$$GP\% = \frac{T_G - T_C}{T_C} \times 100 \quad (5)$$

Both $\lambda$ and GP% decrease with increasing magnetic field; $\lambda$ (GP%) was found to be 0.29, 0.24 and 0.12 (15%, 13% and 12%) for applied fields of 0.01 T, 0.05 T and 0.1 T, respectively in the LS compound.

In the $La_{(1-x)}Sr_xMnO_3$ system, Jahn-Teller (JT) distortions have been identified as the reason for the appearance of the GP[7]. JT distortions exist for $Mn^{3+}$ ions, as there is only one electron in a degenerate $e_g$-state and to reduce its energy there will be a geometrical distortion along one of the fourfold axes. For the $Mn^{4+}$ ion there is no JT distortion since no electron occupies the $e_g$-state. Similarly, $Cu^{2+}$ ions exhibit a JT distortion, while $Cu^{3+}$ ions don't. From the balance of valence charges, $Cu^{2+}$ substitution will increase the amount of $Mn^{4+}$ ions and decrease the amount of $Mn^{3+}$ ions. The substitution is straight forward for $Cu^{3+}$, it only replaces $Mn^{3+}$ ions. As a combined effect of $Cu^{2+}$ and $Cu^{3+}$ substitutions, the ratio of JT ions ($Mn^{3+}$ and $Cu^{2+}$) to non-JT ions ($Mn^{4+}$ and $Cu^{3+}$) will decrease. A direct evidence of this argument is observed in the XPS analysis.

Two types of JT distortions have been identified; pairs of weakly distorted $Mn^{+3}$-$Mn^{+4}$ ions, each pair sharing an electron-hole pair, and isolated $Mn^{+3}$ ions exhibiting considerably larger lattice distortions [37]. The energy barrier for activated hopping of charge carriers will be large for isolated $Mn^{+3}$ ions, while the barrier will be much reduced for pairs of $Mn^{+3}$-$Mn^{+4}$ ions, referred to as dimers by Kumar et al.[38] and dimerons by Downward et al. [39] Dimerons form at some temperature above $T_C$ (here related to $T_G$), where there is a deviation from the Curie-Weiss law in the temperature dependent susceptibility plot, and favour ferromagnetic double-exchange interaction via mobile charge carriers. As the magnetic transition temperature is approached, dimerons will form small clusters and the evolution of these clusters is similar to diffusion limited aggregation [39]. The formation of spin clusters depends on the amount of local lattice distortions and the availability of electron hole-pairs in the compound. The overall decrease of JT-active ions in the Cu substituted compound restricts the formation of spin clusters, implying that the compound from a magnetic perspective becomes more homogeneous.

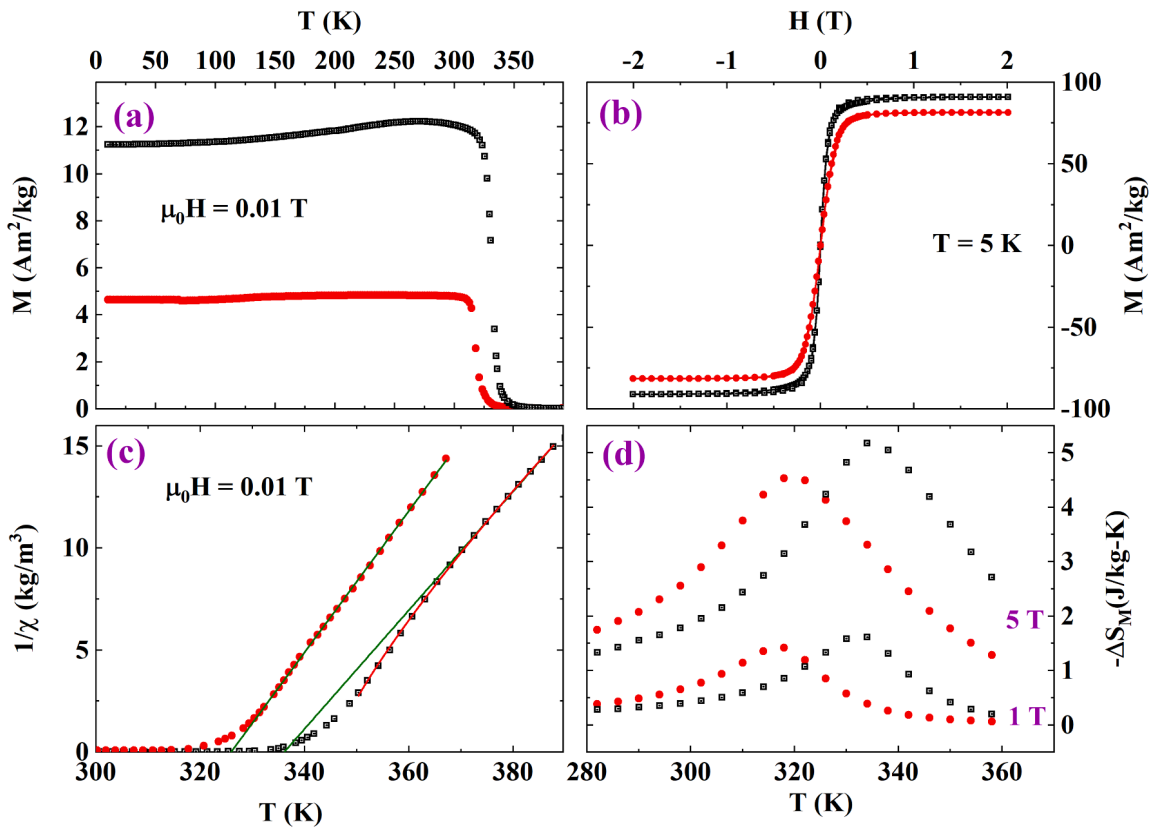

**Fig. 4.** (a) Low-field magnetization versus temperature and (b) magnetization versus magnetic field at 5 K. Temperature dependence of (c) inverse susceptibility and (d) isothermal entropy change for LS and LSC samples. Black (red) data-points give the result for LS (LSC) compound.

The magnetocaloric (MCE) properties have been characterized in terms of the isothermal entropy change using magnetometry. From Maxwell's relation the isothermal entropy change ($\Delta S_M$) can be expressed as,[14]

$$\Delta S_M = \mu_0 \int_{H_i}^{H_f} \left(\frac{\partial M}{\partial T}\right)_H dH \tag{6}$$

where $\mu_0$ is the free-space permeability, $H_i$ and $H_f$ are the initial and final applied magnetic fields, respectively. Also, for real applications it is important to realise a sufficiently large effective temperature range in which a refrigeration process will work and the relative cooling power (RCP)[14] is a measure of this. The RCP is defined as,

$$RCP = -\Delta S_M^{max} \times \Delta T_{FWHM} \tag{7}$$

where $-\Delta S_M^{max}$ is the isothermal entropy change maximum and $\Delta T_{FWHM}$ is the full width at half maximum of the $-\Delta S_M$ versus temperature curve. The field and temperature dependence of the isothermal entropy change for the LS and LSC compounds are shown in Fig. 4(d). The maximum value of isothermal entropy change was observed near $T_C$ for both compounds. At an applied field of 5 T, the entropy maximum and the RCP value decrease by 12.5% and 6%, respectively with 10% Cu-substitution in the B-site (see Table 2). This is also expected from the decrease of the saturation magnetization with Cu-substitution. However, Cu-substitution is still of value as it lowered the temperature where $\Delta S_M^{max}$ occurs towards room temperature.

## 4. Discussion

The effect of partial substitution of Mn with Cu in $La_{0.8}Sr_{0.2}MnO_3$ on structural, electronic, chemical, magnetic and magnetocaloric properties are described here. Rhombohedral (R-3c) structures with almost the same lattice parameters were observed for the two compounds, which allows for a purer investigation of how lattice distortion and competing magnetic interactions (arising from additional AFM interactions of type $Mn^{+3}$-$O^{2-}$-$Cu^{+2}$) affect the evolution of a GP. From XPS analysis negligible amount of O-deficiency was observed for the two compounds, which is also important to avoid effect of anions on the valence charge of B-site magnetic ions. In the LS compound, the GP was observed due to local lattice distortions and aggregation of dimerons, associated with the JT-effect. This spin disordered phase is suppressed in the LSC compound owing to the decrease of the number of JT-active ions. In the LSC compound the ratio of JT/non-JT ions is 2.49, which is close to the value 2.33 observed for $La_{0.7}Sr_{0.3}MnO_3$ compound for which there is also no GP-singularity [40]. However, the $La_{0.7}Sr_{0.3}MnO_3$ compound has a higher $T_C$ (368.45 K), which can be explained by the reduction of the electronic bandwidth with Cu-substitution. Recently, A. Chanda *et al.* [41] have reported a change of $T_C$ of ~ 50 K with a 10% Ga-substitution for Mn in the $La_{0.6}Sr_{0.4}MnO_3$ compound. This change of $T_C$ is five times larger than the decrease of $T_C$ in this work, although in both cases there is a decrease of JT-active ions. Thus, the value of $T_C$ mostly depends on the electronic bandwidth, while the GP-disorder depends on the number of JT-active ions. The Cu-substitution plays two crucial roles, suppressing the disordered GP by reducing the number of JT-active ions and tuning the value of $T_C$ towards room temperature by decreasing the electronic bandwidth. Apart from this, a reasonable value of the relative cooling power and isothermal entropy change near room temperature make B-site substitution with Cu-ions interesting for solid state cooling devices.

**Declaration of Competing Interest**

The authors declare that they have no known competing financial interests or personal relationships that could have appeared to influence the work reported in this paper.

**Acknowledgements**

The Swedish Foundation for Strategic Research (SSF, contract EM-16-0039) supporting research on materials for energy applications is gratefully acknowledged. Infrastructural grants by VR-RFI (#2017-00646_9) and SSF (contract RIF14-0053) supporting accelerator operation are gratefully acknowledged. Financial support by FITC HF RAS through project No. 45.22 (grant AAAA18-118012390045-2) is gratefully acknowledged.